\documentclass[pre,twocolumn,a4paper,superscriptaddress]{revtex4}
\usepackage{amsmath}
\usepackage{amsfonts}
\usepackage{amssymb}
\usepackage{epsfig}
\begin{document}

\title{Culture-area relation in Axelrod's model for culture dissemination}

\author{Lauro A. Barbosa}
\email{lau@ifsc.usp.br}
\affiliation{Instituto de F\'{\i}sica de S\~ao Carlos,
  Universidade de S\~ao Paulo,
  Caixa Postal 369, 13560-970 S\~ao Carlos, S\~ao Paulo, Brazil}   
                      
 \author{Jos\'e F. Fontanari}
\email{fontanari@ifsc.usp.br}
\affiliation{Instituto de F\'{\i}sica de S\~ao Carlos,
  Universidade de S\~ao Paulo,
  Caixa Postal 369, 13560-970 S\~ao Carlos, S\~ao Paulo, Brazil}

\begin{abstract}
Axelrod's model for culture dissemination offers a nontrivial answer to the question of
why there is cultural diversity given that people's beliefs 
have a tendency to become more similar to each other's as
people interact repeatedly. The answer depends on the two  control parameters of the 
model, namely,  the number $F$ of cultural features that characterize each agent, and the number $q$ of traits  that
each  feature can take on, as well as on the  size $A$ of the territory  or, equivalently, on the number of 
interacting agents. Here we investigate the dependence of the number $C$  of distinct coexisting  cultures 
on the area $A$ in Axelrod's model -- the culture-area relationship --  through extensive Monte Carlo simulations.
We find a non-monotonous culture-area relation, for which  the number of cultures decreases when the area grows 
beyond a certain size,  provided that $q$ is smaller than a threshold value $q_c = q_c \left ( F \right )$ and
$F \geq 3$. In the limit of infinite area, this threshold value signals the onset of a 
discontinuous transition between a globalized  regime marked by a uniform culture ($C=1$), and a completely 
polarized regime where all $C = q^F$ possible cultures coexist.
Otherwise the culture-area relation exhibits
the  typical behavior of the  species-area relation, i.e., a monotonically increasing curve the slope of which is steep at first
and steadily levels off at some maximum diversity value.
\end{abstract}


\maketitle

\section{Introduction}\label{sec:Intro}

Axelrod's model for the dissemination of culture or social influence \citep{Axelrod_97} is a 
paradigm for idealized models of
collective behavior which seek to boil down a collective phenomenon  to its functional essence \citep{Goldstone_05}. 
The main issue  the model addresses is why 
cultural differences persist despite the fact that interactions between people tend to make
them more alike in their beliefs and attitudes. Building on just a few simple principles, Axelrod's model  provides a
highly nontrivial answer to that question. In that model, an agent  is represented by a string of 
cultural features, where each feature can adopt a certain number of distinct traits. The interaction between any two agents takes place
with probability proportional to their cultural similarity, i.e., proportional to the number of traits they have in common.
The result of such interaction is the increase of the similarity between the two agents, as one of them modifies 
a previously  distinct trait to match that of its partner.

A remarkable aspect of Axelrod's model is that, notwithstanding the
built-in assumption that social actors have a tendency to become similar to each other through
local interactions, the model exhibits global polarization, i.e., a stable multicultural regime
\citep{Axelrod_97}. Subsequent analysis of this model by the statistical physics community has revealed a rich 
dynamic behavior with a non\-equi\-li\-brium  phase transition separating the global polarization regime
from the homogeneous regime, where a single culture dominates the entire population 
\citep{Castellano_00, Klemm_03a,Klemm_03b,Miguel_05}. An important outcome of those more quantitative
studies was the finding that the multicultural regime is unstable to a vanishingly small
noise that allows for the agents to spontaneously change their opinions  \citep{Klemm_03c}
(see, however, \cite{Parisi_03}).
Several studies of a more qualitative character have considered generalizations of the original
model such as variability in the agents' range of communication and mass media effects
(see, e.g., \cite{Kennedy_98,Shibanai_01,Greig_02}).  These efforts seem to have established  Axelrod's model
as the reference minimal model of social influence or culture dissemination both in the social
and physical sciences \citep{Miguel_05,Toral_07}. We must note, however, that there are many alternative models of social
influence or opinion formation 
which, similarly to Axelrod's, focus on the interplay between consensus and diversity, and which have also been extensively
studied by the statistical physics community (see, e.g., 
\cite{Lewenstein_92,Sznajd_00,Deffuant_00,Galam_02}).

Despite all the interest raised by Axelrod's model, a most appealing outcome of the model -- 
the existence of a multicultural regime -- has been somewhat overlooked and even obvious questions
such as the relation between the number of coexisting cultures and the area available to the
social agents, i.e., the culture-area relation has not been fully addressed. This is 
surprising in view of the counterintuitive result found by \cite{Axelrod_97} that the number of 
coexisting cultures decreases when the area grows beyond a certain size, which starkly contrasts 
with the biological species-area relations characterized by the monotonical increase of the
number of species with the size of the area of a particular habitat
\citep{Rosenzweig_95,He_96,Rosenzweig_99,Stauffer_07}.

Axelrod's model is characterized by two integer-valued parameters, namely,
the list of features or dimensions of culture $F$ and the number of traits $q$ which are the
possible values each feature can take on. The social agents live in the sites of a square lattice of
linear size $L$ and can interact with their  nearest neighbors only.
In this contribution we re-examine the culture-area relation in Axelrod's model and show that the
unusual non-monotonic behavior occurs only in the regime of $F \geq 3$ and $q < q_c$ where 
$q_c = q_c \left ( F \right )$ is the number of traits at which the discontinuous transition takes place
in the limit $L \to \infty$. Otherwise, the culture-area relation exhibits the typical behavior of the 
species-area relation -- a monotonically increasing curve the slope of which is steep at first
and steadily levels off at the maximum diversity value $q^F$. Since there is no `intuitive' reason for the
discontinuous transition to take place at a particular value of $q$ (or to not occur for
$F=2$, for instance), we think  the non-monotonical behavior of 
the culture-area relation has no first-principle explanation, as sought by \cite{Axelrod_97}
in his original work.

This study  is organized as follows: In Sect. \ref{sec:Model} we describe briefly Axelrod's model for
culture dissemination and in Sect. \ref{sec:Simulations} we present and discuss the culture-area relations
obtained from extensive simulations of the two representative cases $F=2$ and $F=3$ for which
the transition between the multicultural and homogeneous regimes is continuous and discontinuous,
respectively. Finally,
in Sect. \ref{sec:Conclusion} we relate our findings to results of models of language competition
\citep{Stauffer_05,Schulze_08} and discuss a possible connection between Axelrod's model and the Derrida-Higgs
model for sympatric speciation \citep{Higgs_91,Higgs_92,Manzo_94}.

\section{Model}\label{sec:Model}

As pointed out before, in Axelrod's model each agent is characterized by a set of $F$ cultural features which can take on
$q$ distinct values. The agents are fixed in the sites of a square lattice with open boundary conditions
(i.e., agents in the corners interact with two neighbors, agents in the sides  with three, 
and agents in the bulk with four nearest neighbors). The social
agents  can be thought of  as  individuals or as homogeneous villages.
The initial configuration is completely random with the features of each agent given by  
random integers drawn uniformly between $1$ and $q$. At each time we pick an agent at random
(this is the target agent) as well as one of its neighbors. These two 
agents interact with probability equal to  their cultural similarity, defined as the fraction of 
common cultural features. An interaction consists of selecting at random one of the distinct features, and changing the
target agent's trait on this feature to the neighbor's corresponding trait. This procedure is repeated until 
the system is frozen in an absorbing configuration. 

Given the bias towards homogenization, it is really remarkable that in some cases
the system can reach a multicultural  absorbing state. We recall that the sources of
disorder in Axelrod's model are the stochastic update sequence and the choice of the initial configuration: 
it is the competition between the disorder of the initial configuration and  the ordering bias of the local  interactions that 
is responsible for the nontrivial threshold phenomenon reported  by \cite{Castellano_00}.

\section{Simulations}\label{sec:Simulations}

A feature that sets our results apart from those reported previously in the literature is that
our data points represent averages over at least $10^3$ independent runs. This requires a substantial
computational effort, especially in the regime where the number of cultures decreases with
the lattice size since then the time for absorption can be as large as $10^6 \times A$ where
$A = L^2$ is the lattice area. In the figures presented in the following,
the error bars are smaller or at most equal to the symbol sizes.

To simulate efficiently Axelrod's model we  make a list of the active agents. An active agent is an agent 
that has at least one feature in common  and at least one feature distinct with at least one of its nearest neighbors. 
Clearly, since only active agents can change their cultural features, it is more efficient  to select the
target agent randomly from the list of active agents rather than from the entire lattice. Note that the
randomly selected neighbor of the target agent may not necessarily be an active agent itself. In the case that
the cultural features of the target agent are modified by the interaction with its neighbor, we 
need to re-examine the active/inactive status of the target agent as well as of all its neighbors so as
to update the list of active agents.  The dynamics is frozen when the list of active agents is empty.
 
Our focus is on the number of distinct cultures $C$, rather than on the number of clusters or
the fraction of the lattice occupied by the largest cluster \citep{Castellano_00, Klemm_03a,Klemm_03b}. In that sense,
cultural diasporas, which occurs when regions with specific cultural features are disconnected from other
regions with the same cultural features \citep{Greig_02}, are counted as a single culture.
 Of course, $C$ is much easier to compute than the cultural regions and, as we will show next, provides an 
equally good indication of the existence and location of a threshold phenomenon.
In addition, we consider the cases $F=2$ and $F=3$ only as, according to \cite{Castellano_00}, the dynamic and static properties 
of Axelrod's model for $F > 3$ are qualitatively similar to those for $F=3$.

\begin{figure}
\centerline{\epsfig{width=0.47\textwidth,file=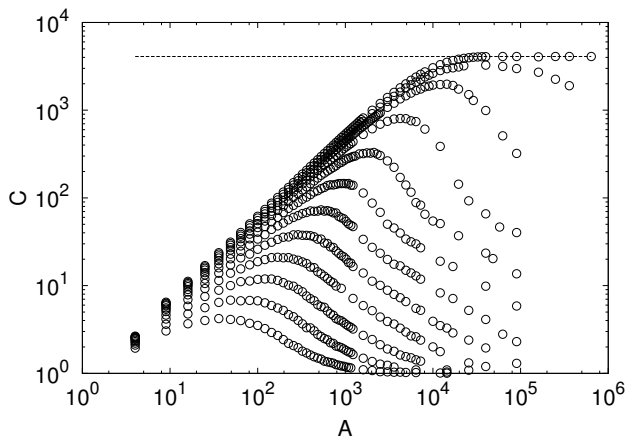}}
\par
\caption{Logarithmic plot of the culture-area relation for $F=3$ and (bottom to top) $q=5,6,\ldots,16$.
The dashed horizontal line indicates the value $C = 16^3$. In the limit $A \to \infty$ 
there are two distinct regimes: either $C \to 1$  ($q < 16$) or $C \to q^F$ ($q \geq 16$).
}
\label{fig:1}
\end{figure}

Figure \ref{fig:1} exhibits the culture-area relation for $F=3$ and different values of $q$. The non-monotonic
behavior reported by \cite{Axelrod_97} appears for $q < q_c = 16$ only. For $ q \ge q_c $ the number of cultures
increases linearly with increasing $A$ at first and then gradually flattens when $A$ becomes on the
order of the maximum value, $q^F$. In the limit $L \to  \infty$ we have only two possible outcomes:
if $q < q_c$ then $C \to 1$ and a  single culture dominates the lattice (ordered regime), otherwise $C \to q^F$ and
all cultures are represented in the lattice (disordered regime). The transition between these two regimes is discontinuous
because $C$ jumps from $1$ to $q^F$ at $q=q_c$. As mentioned before, this behavior is expected to
occur for all $F \geq 3$ with the threshold value $q_c = q_c \left ( F \right )$ increasing monotonically
with increasing $F$ \citep{Castellano_00}.

\begin{figure}
\centerline{\epsfig{width=0.47\textwidth,file=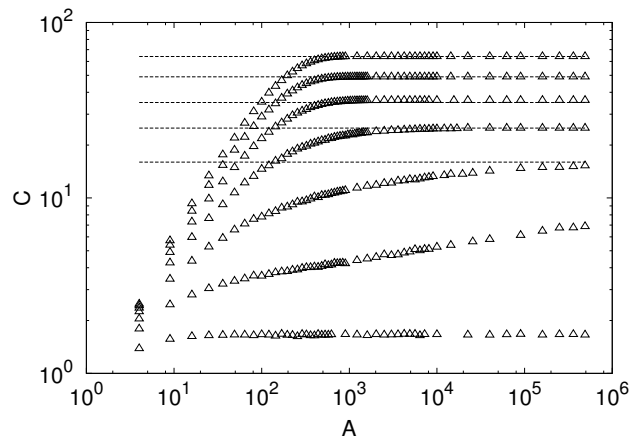}}
\par
\caption{Logarithmic plot of the culture-area relation for $F=2$ and (bottom to top) $q=2,3,\ldots,8$.
The dashed horizontal lines indicate the  maximum diversity values $q^F$ to which the data for $q \geq 4$
converge in the limit of infinite area. 
}
\label{fig:2}
\end{figure}

Figure \ref{fig:2} summarizes our findings regarding the case $F=2$. The first point to be noted is that,
in contrast with the previous case, the culture-area relation exhibits the expected monotonic behavior, which
implies that the globally homogeneous regime $C=1$ does not appear  in the limit $A \to \infty$. But the disordered
regime, which is characterized by the coexistence of all $q^F$ cultures, 
is present as revealed by the data for $q \geq 4$.
We can identify a second regime (see data for $q = 2$), in which only a fraction of the total number of cultures coexist in the limit of
infinite lattices. In this limit we find $ C \to 1.66 \pm 0.01$ for $q=2$. It is not clear whether the data for 
$q=3$ will ultimately converge to $C=9$: for $L=700$  we find $ C = 6.91 \pm 0.15$ but the data  show
a trend to increase much further.
The very slow convergence may indicate that $q=3$ is the threshold (critical)
value that separates the two regimes. In the case  $q$ is allowed to change continuously (for example, by choosing the trait values
as samples of a Poisson distribution of mean $q$), \cite{Castellano_00} have shown that the transition  between these two 
regimes is  continuous, in the sense that, for $A \to \infty$, $C$   increases continuously from $1$ to $q_c^2$ as the mean of
the Poisson distribution  varies from $0$ to $q_c$.

\begin{figure}
\centerline{\epsfig{width=0.47\textwidth,file=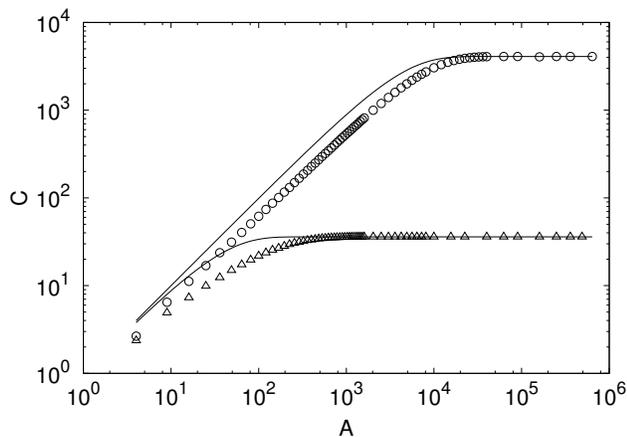}}
\par
\caption{Comparison between the random initial diversity $C_r$ (solid lines) given by 
Eq.\ (\ref{Er}) and the stationary diversity of Axelrod's model for
$F=3$, $q=16$ ($\circ$), and   $F=2$, $q=6$ ($\triangle$). 
}
\label{fig:3}
\end{figure}

It is instructive to calculate the number of cultures in the totally disordered initial configuration, in
which the $A = L^2$ agents are assigned  one of the any $q^F$ cultures. This  
is a classical occupancy problem discussed at length in Feller's book \cite[Ch. IV.2]{Feller:68}.
In this occupancy problem, the probability that exactly $m$ cultures are not used in the assignment of the $A$ agents to 
the $q^F$ cultures is 
\begin{equation}\label{m}
P_m \left (A, q^F \right ) = \left ( \begin{array}{c} q^F \\ m \end{array} \right ) 
\sum_{\nu =0}^{q^F-m} \left ( \begin{array}{c} q^F-m \\ \nu \end{array} \right ) \left ( -1 \right )^{\nu} 
\left ( 1 - \frac{m + \nu}{q^F} \right )^A ,
\end{equation}
which in the limit where  $A$ and $q^F$ are large reduces to the Poisson distribution 
\begin{equation}\label{poisson}
p \left (m; \lambda \right ) = \mbox{e}^{-\lambda} \frac{\lambda^m}{m!}
\end{equation}
where $\lambda = q^F \exp \left ( - A/q^F \right )$ remains bounded \cite[Ch. IV.2]{Feller:68}. Hence the average 
cultural diversity $C_r$ resulting from the random assignment of agents to cultures is simply 
$ q^F -  \left \langle m \right \rangle$, which yields 
\begin{equation}\label{Er}
C_r = q^F \left [ 1 - \exp \left ( - A/q^F \right ) \right ] .
\end{equation}
This quantity is a monotonically increasing function of $A$ which grows linearly 
in the regime  $A \ll q^F$ and tends to the maximum diversity value  $q^F$ when $A \gg q^F$. Figure
\ref{fig:3} shows a comparison between the predictions of Eq.\ (\ref{Er}) and the simulation data
of Axelrod's model in the global polarization (multicultural)  regime. Although the random
occupancy hypothesis yields a good  qualitative description of the culture-area relations in this
regime, it  consistently overestimates the  values for the cultural diversity. This is  expected
as the effect of the local interactions in Axelrod's model is to decrease the cultural differences
between neighboring agents.

\section{Conclusion}\label{sec:Conclusion}

In contrast with the species-area relation of Biology for which there are plenty
of field data to check the theoretical proposals \citep{Rosenzweig_95}, in
the culture-area relation there are practically no empirical evidences to back
any quantitative theoretical prediction. However, there are some empirical 
results regarding the language-area relation \citep{Nettle_98,Nettle_99}, which
are appropriate to mention here since the mechanisms of development, dissemination and
acquisition of language are similar, if not identical, to those of culture. An
extensive analysis of the language diversity  that considers ecological and linguistic
variables  for about $74$ countries  yields  $C \propto A^x$ with $x = 0.5 \pm 0.1$ \citep{Nettle_98}.
In the  region $q^F \gg A$, Axelrod's model yields also a power-law scaling but with the exponent $x=1$.
Given the crudeness of the model and the inherent difficulties involved in counting languages, either
agreement or disagreement on this matter seems to be of little significance. Nevertheless, it should
be interesting to find out whether changes in  rules for the local interaction between agents can
affect the value of that exponent.
We note that the area $A$ used in the field studies is the area of the country, whereas 
in Axelrod's model $A$ is the area of the lattice or, equivalently, the number of agents (population size). 
The exponent $x$ is not affected by these different interpretations of $A$, provided
the population size grows linearly with the territory area.

Interestingly, extensive Monte Carlo simulations of a language competition model \citep{Stauffer_05,Schulze_08} 
yield a non-monotonic  relation between the number of languages and the number of speakers (agents). 
In fact, the similarity between the language-area (population size) relation obtained in the case the individuals 
are placed in the sites of a scale-free network (see Fig. 4 of \cite{Schulze_08}) and  the relations shown in Fig.\ 
\ref{fig:1} for $q < 16 $ is striking. The model for language competition proposed by \cite{Schulze_08} has
an important  element in common with Axelrod's model --  a language is 
defined by $F$ independent features each of which can take 
one of $q$ different values. The local interaction rules, however, are completely distinct and, for instance,
the similarity between the agents' languages has no role in determining the occurrence of an interaction.
Unless there is an explicit dependence of the usefulness of a language on spatial coordinates \citep{Patriarca_04},
the  ultimate outcome of the language competition models is the  dominance of a single language \citep{Abrams_03} 
(see, however, \cite{Schulze_08} for an overview of language models which may exhibit  language coexistence)
and so
since the source of diversity or randomness is identical in both models  (the initial distribution of languages
and cultures) the similarities  pointed out  may not be so surprising after all.

A word is in order about the relevance of Axelrod's model for theoretical biology and, in particular,
for the description of the dynamics of species in a habitat. From the broad perspective, we must point out that
the case for the pertinence
of culture to  the understanding of the  hominid  evolutionary  process has  been persuasively  advocated
by \cite{Laland_06} using the biological concept of niche construction \citep{Laland_99}. More parochially,
however, we can interpret  Axelrod's model as a  model for sympatric speciation based on a mate choice  mechanism
that depends on the similarity (genetic distance) between mates (see \cite{Schluter_00} for a review of several
mechanisms by which sexual selection can drive speciation). One such model is the Derrida-Higgs model of species
formation \citep{Higgs_91,Higgs_92,Manzo_94} in which mating only occurs between individuals that are genetically
similar to each other. This assumption is akin to the restriction  of the interactions in Axelrod's model
to individuals that share a certain number of cultural traits. In addition, we can re-interpret cultural 
assimilation by the target agent as the result of a sexual reproduction scheme implemented  
by a Moran-type stochastic process in which the
target agent is  replaced by the offspring whose genotype is identical to the target's genotype except for
a single gene which is inherited from the other mate. Although a more traditional crossover scheme is unlikely to
change qualitatively the outcome of the competition between dominance and diversity 
discussed in this paper,  it would be interesting to study the Derrida-Higgs model of species formation
with the individuals fixed in the lattice sites and the mating restricted to their nearest neighbors as in 
Axelrod's model.

The paucity of empirical data to support and motivate the proposal of models for
culture dissemination and social influence is about to change as more people become
connected by the Web 2.0 social networks. The on-line communities in these 
networks can provide an invaluable source of data to validate theoretical predictions
of models such as Axelrod's.
In fact, the basic idea that agents who are similar to each other are more likely to interact
(`birds of a feather flock together') and then become even more similar was observed in that
context by \cite{Singla_08}. Analysis of a population of over $10^7$ people indicates 
that people who chat with each other using instant messaging are more likely to have common interests,
as measured by the similarity of  their Web searches, and the more time they spend talking,
the stronger this relationship is.

\begin{acknowledgements}
The  work of J.F.F. was supported in part by CNPq and FA\-PESP, Project No. 04/06156-3. L.A.B. was supported by a 
FAPESP postdoctoral fellowship.
\end{acknowledgements}

\end{document}